\documentstyle[preprint,%
aps]{revtex}

\begin{document}
\draft
\preprint{hep-ph/9308232; BI-TP 93/42}
\title{The Nonabelian Debye Mass 
at Next-to-Leading Order}
\author{A. K. Rebhan\cite{byline}}
\address{Fakult\"at f\"ur Physik, Universit\"at Bielefeld,\\
D--33501 Bielefeld, Germany}
\maketitle
\begin{abstract}
It is shown that after a resummation of leading high-temperature
contributions, a complete and gauge-independent result for
the nonabelian Debye screening mass at next-to-leading order can
be extracted from the static gluon propagator. In contrast to
previous, incomplete results, the correction to the Debye
mass is found to be logarithmically sensitive to the
nonperturbative magnetic mass and positive, in accordance
with recent high-statistics results from lattice calculations.
\end{abstract}

\narrowtext
\tighten

Over the last few years it has become clear that a complete
calculation of perturbative
corrections to the dispersion laws of quasi-particles in
high-temperature QCD requires a resummation of
the leading-order terms with characteristic scale $gT$,
where $g$ is the coupling constant and $T$ the temperature.
An improved perturbation theory has been developed in
particular by Braaten and Pisarski \cite{BP} and has been
applied in the first place to derive damping effects of
collective excitations \cite{Damping}, where it turned
out to be indispensable in order to obtain accurate and
gauge independent results for the leading-order term $\sim g^2T$.
Most recently it has also been employed successfully in a
calculation of the next-to-leading order term of the
QCD plasma frequency $\delta\omega^2_{\rm pl} \sim g \omega^2_{\rm pl}$
\cite{HS}.

All these applications are still within the limits of
a perturbative barrier set by the presumably incalculable screening
of static magnetic fields \cite{Linde}, which causes a
breakdown of perturbation theory at a certain order. In the
case of the gluon self-energy from which corrections to dispersion
laws are extracted this incalculability is expected at the
order $g^4T^2$, whereas the above-mentioned results are
derived from this quantity evaluated up to order $g^3T^2$.
However, even at this order occasionally a sensitivity to
the behaviour of static transverse gluons at momentum scale $\ll gT$
has been found, giving rise to a logarithmic enhancement
$
\sim \ln(1/g)$.

Besides the dispersion law of propagating collective modes, 
a quantity of singular importance
which can be derived from the gauge boson self-energy is the
static (chromo)electric screening (Debye) mass \cite{Kapusta}.
At leading order $\Pi_{00}(k_0=0,{\bf k})=m_0^2\sim g^2T^2$,
which gives rise to a pole in the
static gluon propagator at ${\bf k}^2=-m_0^2$, and, consequently,
to exponential screening in the potential
\begin{eqnarray}
&&\Phi(r)= Q\int\frac{d^3k}{(2\pi)^3}
\frac{e^{i{\bf k}{\bf r}}}{k^2+\Pi_{00}(k_0=0,k)}\nonumber\\
&=&\frac{Q}{(2\pi)^2}\int_{-\infty}^\infty
\frac{sin(kr)}r \frac{k\,dk}{k^2+\Pi_{00}(0,k)}
=\frac{Q}{4\pi r}e^{-m_0r}
\label{v}\end{eqnarray}
with
\begin{equation}
m_0^2=\frac{g^2 (N+N_f/2)T^2}3,
\end{equation}
for gauge group SU($N$) and $N_f$ fermions.
In linear response theory, the gradient of $\Phi$ gives the
longitudinal electric field generated by a static charge $Q$.

Again, one
would expect perturbative calculability of the screening mass
at the next-to-leading order
$\delta \equiv \delta m^2/m_0^2 \sim g$.
Early attempts (using temporal gauge)
\cite{KK1,KK2} gave $\delta = - c g$, with $c$ a positive constant
depending on the particular partial resummation employed, whereas
in general axial gauge even a different sign was obtained \cite{FK}.
In covariant gauges, Toimela \cite{T} found that the time-time
component of the gluon self-energy
$\Pi_{00}(k_0=0,{\bf k}\to0)$ is gauge dependent at
the order $g^3T^2$, which was confirmed
by Nadkarni \cite{N1} on the basis of a dimensionally
reduced effective theory. 
This was taken to mean \cite{N1} that the Debye
mass could not be extracted from the electrostatic propagator.

However, all these investigations concentrated
on the limit ${\bf k} \to 0$ of the static time-time component $\Pi_{00}$,
guided by the fact that this limit indeed yields the
leading-order contribution to the Debye mass. But
at leading order $g^2T^2$,
$\Pi_{00}(k_0=0,{\bf k})$ happens to be independent
of ${\bf k}$, whereas
the pole of the corrected electrostatic propagator
is located at finite $|{\bf k}|\sim gT$.
Gauge dependence at ${\bf k} \to 0$, away from the physical pole of the
propagator, is thus only to be expected in a nonabelian gauge theory.
Conversely, formal arguments exist that the
relevant poles of a self-consistently
corrected propagator should be gauge-independent also in
the nonabelian theory \cite{KKR}. And it is just the pole
in Eq.~(\ref{v}) that determines the exponential
decay.

Incidentially, in QED it is equally important to define the
electric screening mass in a self-consistent manner by the
location of the pole rather than the ${\bf k}\to0$ limit
of $\Pi_{00}$ as almost invariably done in the literature. This
in fact modifies the QED Debye mass squared at and above the order
$e^4T^2$ \cite{QED}.

According to the resummation program of Ref.~\cite{BP},
a complete calculation of the next-to-leading order term
in the nonabelian Debye mass
should be possible and requires the resummation of
all hard-thermal-loop contributions.
In fact, because only a static quantity
is to be calculated, this task can be greatly simplified. In
the problem of determining next-to-leading order corrections
to the effective potential in thermal field theories,
such a simplified scheme has been employed recently by
Arnold and Espinosa \cite{AE}. They have found that in gauge
theories 
it is algebraically much simpler to resum only static modes.
This does not touch the completeness of the
resummation because, in the imaginary-time formalism,
nonstatic modes always imply $K^2\equiv-((2\pi n T)^2+k^2)
\gtrsim T^2 \gg (gT)^2$,
so that hard-thermal-loop corrections are truly perturbative.
Separating the static modes does however give away the
possibility of a straightforward analytic continuation of
any external frequencies.
The full resummation scheme of Braaten and
Pisarski thus is mandatory when
external frequencies $\sim gT$ are to be considered as in
dynamical properties of quasi-particles.

The next-to-leading order contributions to static quantities
of relative order $O(g)$
are determined by resummation of one-loop diagrams, and
for them
it suffices to take into account only the static modes (which
also precludes fermionic contributions).
A further major simplification arises in that with all gluon
lines being static the hard-thermal-loop corrections to the
vertices vanish.
Only the static gluon propagator is thus needed, which reads
\widetext
\begin{equation}
\Delta_{\mu\nu}\bigg|_{p_0=0}=\left[
\frac1{{\bf p}^2+m_0^2}
\delta^0_\mu \delta^0_\nu
+\frac1{{\bf p}^2}
\left(\eta_{\mu\nu}-\delta^0_\mu \delta^0_\nu +
\frac{P_\mu P_\nu}{{\bf p}^2} \right)
+\alpha \frac{P_\mu P_\nu}{({\bf p}^2)^2} \right]_{p_0=0},
\label{stpr}\end{equation}
where $P=(p_0,{\bf p})$ and $\alpha$ is the gauge parameter of covariant
gauges.

Evaluating $\Pi_{00}$ at relative order $O(g)$ then yields
\begin{eqnarray}
\delta\Pi_{00}(k_0=0,{\bf k})
&=&gmN\sqrt{\frac6{2N+N_f}}\int\frac{d^3p}{(2\pi)^3}
\biggl\{\frac1{{\bf p}^2+m^2}+\frac1{{\bf p}^2} \nonumber\\
&+&\frac{4m^2-({\bf k}^2+m^2)
[3+2{\bf p}{\bf k}/{\bf p}^2]}{{\bf p}^2({\bf q}^2+m^2)}
+\alpha ({\bf k}^2+m^2)
\frac{{\bf p}^2+2{\bf p}{\bf k}}{{\bf p}^4({\bf q}^2+m^2)} \biggr\},
\label{pi00}\end{eqnarray}
where ${\bf q}={\bf p}+{\bf k}$ and dimensional regularization is understood
\cite{AE} (yielding
no pole terms because of the odd integration dimension). Here and
in the following the index 0 on $m$ has been dropped, the difference
in $\delta\Pi$ being formally of higher order.
\narrowtext

This result shows that $\delta\Pi_{00}(k_0,{\bf k}\to0)$ is indeed
gauge dependent, as found in Ref.~\cite{N1} in a one-loop calculation
in the dimensionally reduced effective theory, to which the
above reasoning has in fact boiled down as far as the $O(g)$
correction is concerned. It also shows that this gauge dependence
is to disappear on an algebraic level when going to the imaginary pole
${\bf k}^2=-m^2$.

Closer inspection reveals, however, that with ${\bf k}^2\to-m^2$
there appear "mass-shell" singularities caused by the massless
denominators in Eq.~(\ref{pi00}). A simple introduction of
a magnetic mass in the denominator in front of the second term
of the gluon propagator, Eq.~(\ref{stpr}), provides a
physical cut-off, and because the singularities at hand are
only logarithmic, the coefficient of the ensuing logarithm
should be insensitive to the detailed structure of the infrared
limit of the transverse propagators \cite{BMAR,PPS}. All
the massless denominators figuring in the $\alpha $-independent
part of Eq.~(\ref{pi00}) are indeed
associated with the magnetic sector (as opposed to pure gauge
modes).
Assuming $m_{\rm magn}\sim g^2T$, the leading contribution to
$\delta\Pi_{00}$ is determined by the logarithmically divergent pieces
and reads
\begin{mathletters}\label{ds}
\begin{eqnarray}
&&\delta\Pi_{00}(k_0,{\bf k})\bigg|_{{\bf k}^2=-m^2}\nonumber\\
&\equiv&\delta m^2
=gm^2\frac{N}{2\pi}\sqrt{\frac6{2N+N_f}}\ln\frac1g+O(g),
\label{dln}\end{eqnarray}
giving an unexpectedly large and
positive correction to the nonabelian Debye mass.
The positive sign seems to be a genuine nonabelian effect,
for the next-to-leading order correction of the Debye mass
in e.g.\ scalar QED is negative \cite{KRS}. Moreover, no logarithmic
enhancement occurs in the latter. [In spinor QED there are no
$O(g)$ corrections at all due to Pauli suppression.]

The sublogarithmic terms of course do depend on the detailed
structure of the infrared limit of the transverse propagators,
and so cannot be determined completely in the present resummed
one-loop calculation. However, adopting the hypothesis that
this infrared limit just amounts to a finite contribution
$-\frac12\Pi_{ii}(k=0,{\bf k}\to0)=m_{\rm magn}^2\sim g^4T^2$
one may go on to estimate these sublogarithmic terms from
Eq.~(\ref{pi00}).

Here one encounters a subtle difficulty with the $\alpha $-dependent
term in Eq.~(\ref{pi00}), because by approaching the imaginary
pole ${\bf k}^2\to-m^2$,
the explicit factor that apparently ensures gauge independence
gets cancelled by a linear singularity in the momentum integral.
Exactly the same phenomenon was encountered in the recalculation
of plasmon damping rates in general covariant gauges in Refs.~\cite{BKS},
while being absent in homogeneous (``strict'') gauges.
In Ref.~\cite{BKSC} I have argued that this behaviour just
reflects a singular, gauge dependent behaviour of the {\it residue} of the
propagator rather than an actual gauge dependence of the pole
determining the dispersion laws. Indeed, introducing an (unphysical)
cut-off again moves the gauge dependence seemingly afflicting
the pole position into the residue, while the correction to the
pole position becomes independent of this infrared regularization.
In this way, the sublogarithmic terms are determined \cite{SC}
and give rise to
\begin{equation}
\delta m^2
=gm^2N\sqrt{\frac6{2N+N_f}}\frac1{2\pi}\left(\ln\frac{2m}{m_{\rm magn}}
-\frac12\right)+O(g^2).
\label{d}\end{equation}
\end{mathletters}

Quite recently, lattice simulations have been performed which
permit the extraction of corrections to the classical Debye
screening length. A rather high precision has been reached in
simulations of pure SU(2) gauge theory at temperatures up to
nearly 8 times the critical temperature \cite{ILMPR}, with the
finding of a {\it positive} excess in Debye mass squared of
$\delta=+0.30(9)$ at $\beta=3$, corresponding to $T\approx 7.8 T_c$
and $g_R^2\approx 1.19$. Unfortunately, the coupling 
is rather large so that a quantitative comparison with just
the logarithmically enhanced result of Eq.~(\ref{dln})
is out of question. However, taking the result of Eq.~(\ref{d})
seriously and inserting a value for the magnetic mass as suggested
by lattice simulations \cite{MM}, $m_{\rm magn}\approx
0.1365 g^2T$ for SU(2), Eq.~(\ref{d}) yields $\delta=+0.51$, which
comes remarkably close considering the largeness of $g$.

A significantly increased Debye mass has been found previously
also in lattice
simulations of pure SU(3) \cite{SU3}, however with larger statistical
errors.

It should be noted that the previous, incomplete results for corrections
to the nonabelian Debye mass \cite{KK1,KK2,FK,T} have mostly yielded
a negative value for $\delta$, albeit mutually disagreeing in
numerical magnitude. The complete gauge-independent result presented here
differs from the former in that all relevant hard-thermal-loop
contributions are identified and resummed,
and also in that the Debye mass is
defined through the pole of the electrostatic propagator rather
than the (gauge-dependent) zero-momentum limit of the gluon self-energy.

It is the pole that determines the exponential screening
in Eq.~(\ref{v}), whereas the pre-exponential factor therein
will generally be gauge dependent. At the next-to-leading order
considered here, gauge-dependent contributions to the latter
indeed arise from the
residue of the pole of the propagator as well as from a logarithmic
branch cut contribution, which even gives rise to a (gauge-dependent)
modification of the pre-exponential $1/r$-behaviour.

\acknowledgments
I am grateful to R. Baier, J. Kapusta,
U. Kraemmer, and H. Schulz for useful
discussions, and to D. Miller and T. Reisz
for explaining to me the findings
of Ref.~\cite{ILMPR}.

\appendix
\section*{}

In this appendix I give the complete contributions to
$\Pi_{\mu\nu}(k_0=0,{\bf k})$ at order $g^3T^2$ for $k\gtrsim m\sim gT$,
in general covariant gauge. The static limit of $\Pi_{\mu\nu}$ is
transverse, so there are exactly two independent structure
functions, $\Pi_{00}$ and $\Pi_{ii}$.

The correction to $\Pi_{00}$ as given by Eq.~(\ref{pi00}) is
found to be
\begin{equation}
\delta\Pi_{00}(0,k)=\frac{g^2NmT}{2\pi}\left[
\frac{m^2-k^2}{mk} \arctan\frac{k}m+\frac{\alpha -2}2 \right]
\label{pi00a}\end{equation}
for $k^2>0$. This can be immediately continued analyticly to $k^2<0$,
but for $-k^2\equiv \kappa^2 \to m^2$, it has a logarithmic singularity.
As argued in the text, the massless denominators in Eq.~(\ref{pi00})
require an infrared cut-off. Let this for simplicity be
a mass term $\lambda \ll m$
uniformly for transverse and gauge modes. In the case of
the transverse gluons this can eventually be identified with
the magnetic screening mass, whereas in the case of the massless
gauge modes it is just a regulator which drops out from the end
results, Eqs.~(\ref{ds}). Then for $\kappa\approx m$ Eq.~(\ref{pi00a})
gets replaced by
\widetext
\begin{eqnarray}
\delta\Pi_{00}(0,k)&=&\frac{g^2NmT}{4\pi}\biggl\{
\frac{m^2+\kappa^2}{m\kappa}
\ln\frac{(m+\kappa)^2-\lambda^2}{m^2-(\kappa-\lambda)^2}-1
\nonumber\\
&&-(1-\alpha )(m^2-\kappa^2)
\frac1{(m-\lambda)^2-\kappa^2}
\biggr\}.
\end{eqnarray}
\narrowtext
In the limit $\kappa\to m$ this gives the gauge-independent result
for the Debye mass discussed in the text. For general $k$ it
remains gauge-dependent, however, so that beyond leading order
there is no gauge invariant meaning to be attributed to the
dielectricity $\epsilon=1-\Pi_{00}(0,k)/k^2$
as defined from the gluon propagator.

For $k\gg m$,
\begin{equation}
\Pi_{00}(0,k)\to -\frac14 g^2NkT
\end{equation}
becomes independent of the gauge parameter, but in fact would
be different in a background covariant gauge \cite{EHKT}.
This term would give rise to a modification of
the Debye screening in a gluon plasma \cite{GK}, if
such a behaviour persisted for soft
momentum $k\lesssim gT$. However, the complete result of
Eq.~(\ref{pi00a}) reveals that it does not.

The next-to-leading order correction to the
other structure function, $\Pi_{ii}(0,k)$,
can be derived in a similar manner, and is found to be
\begin{eqnarray}
\delta\Pi_{ii}(0,k)&=&g^2NmT\biggl\{
\frac{(\alpha +1)^2+10}{16} \frac{k}m \nonumber\\
&&+\frac1{4\pi}\left[2-\frac{k^2+4m^2}{mk}\arctan\frac{k}{2m}\right]
\biggr\}.
\label{piii}\end{eqnarray}
Again, the magnetic permeability defined by
$1/\mu=1-\frac12\Pi_{ii}/k^2$ is a gauge-dependent quantity
beyond leading order. The gauge-dependent terms vanish
only at the location of the pole of the transverse gluon propagator,
which is at ${\bf k}=0$. There the correction term vanishes
completely,
which means that there is no magnetic mass squared of the
order $g^3T^2$.
The magnetic mass must therefore be $\ll g^{3/2}T$.

For small $k\ll m$, Eq.~(\ref{piii}) has a linear behaviour
with gauge dependent, but positive definite coefficient.
The transverse propagator therefore has the form
$1/(k^2-c k)$ with $c\sim g^2T$. This corresponds to
a pole at space-like momentum with $k\sim g^2T$ \cite{KalKl}. However,
in this regime the perturbatively incalculable contributions
$\sim g^4T^2$ to $\Pi_{\mu\nu}$ become relevant and are expected
to remove this pathology by the generation of a magnetic mass term.


\begin{references}
\bibitem{SC}
Here I am implicitly adopting a self-consistent procedure
in which it is important to replace $m_0\to m$ in Eq.~(\ref{stpr}).
Otherwise, the argument of the logarithm would be modified
and gauge dependent terms would survive in the sublogarithmic
contributions. See however Refs.~\cite{PPS}.
\end{references}
\end{document}